\documentclass{ragtime}
\usepackage{graphicx}

\title[Determination of Characteristics of Eclipsing Binaries with Spots]
      {Determination of Characteristics of Eclipsing Binaries with Spots: Phenomenological vs Physical Models}

\author[Mariia G. Tkachenko, 
        Ivan L. Andronov]
        {Mariia G. Tkachenko\at{a}
        Ivan L. Andronov\at[]{b}\\
 \ins{} Odessa National Maritime University, Mechnikov st. 23, UA-65029 Odessa, Ukraine\\
 \ins{a}\Email{masha.vodn @ yandex. ua}, 
 \ins{b}\Email{tt_ari @ ukr. net}}
\begin{document}
\begin{abstract}We discuss methods for modeling eclipsing binary stars using the "physical", "simplified" and "phenomenological" models..
There are few realizations of the "physical" Wilson-Devinney (1971) code and its improvements, e.g. Binary Maker, Phoebe. A parameter search using the Monte-Carlo method was realized by Zola et al. (2010), which is efficient in expense of too many evaluations of the test function. We compare existing algorithms of minimization of multi-parametric functions and propose to use a "combined" algorithm, depending on if the Hessian matrix is positively determined. To study methods, a simply fast-computed function resembling the "complete" test function for the physical model. Also we adopt a simplified model of an eclipsing binary at a circular orbit assuming spherical components with an uniform brightness distribution. This model resembles more advanced models in a sense of correlated parameter estimates due to a similar topology of the test function. Such a model may be applied to detached Algol-type systems, where the tidal distortion of components is negligible. 
\end{abstract}
\section{Introduction}
Determination of the model parameters of various astrophysical objects, comparison with 
observations and, if needed, further improvement of the model, is one of the main directions of science, 
particularly, of the study of variable stars. And so we try to find the best method for the determination of the parameters of eclipsing binary stars. 
For this purpose, we have used observations of one eclipsing binary system, 
which was analyzed by \citep{Zol:2010}. This star is AM Leonis, which was
observed using 3 filters (B, V, R). For the analysis, we used the computer code written by Professor Stanis{\l}aw
Zo{\l}a \citep{Zol:1997}. In the program, the Monte-Carlo method is
implemented. As a result, the parameters were determined and the corresponding
light curves are presented in the paper \citep{And:2013a}

With an increasing number of evaluations, the points are being concentrated
to smaller and smaller regions. And, finally, the “cloud” should converge to
a single point. Practically this process is very slow. This is why we try to find
more effective algorithms. At the “potential -- potential” diagram \citep{And:2013a}, we see that the best solution corresponds to an “over-contact” system, which makes an addition link of equal potentials $\Omega_1=\Omega_2$ and corresponding decrease of the number of unknown parameters. 

Such a method needs a lot of computation time. We had made fitting using
a hundred thousands sets of model parameters. The best 1500 (user defined) points are stored in the file and one may plot the “parameter – parameter diagrams”. 
Of course, the number of parameters is large, so one may choose many pairs of
parameters. However, some parameters are suggested to be fixed, and thus
a smaller number of parameters is to be determined.

Looking for the “parameter--parameter” diagrams, we see that there are strong
correlations between the parameters. E.g. the temperature in our computations is
fixed for one star. If not, the temperature difference is only slightly dependent on
temperature, thus both temperatures may not be determined accurately from modeling. 
So the best solution may not be unique; it may fill some sub-space in the
space of parameters. 

This is a common problem: the parameter estimates are dependent. Our tests
were made on another function, which is similar in behavior to a test function
used for modeling of eclipsing binaries. 

To determine the statistically best sets of the parameters, there are some methods for optimization of the test function which is dependent on these parameters \citep{Cher:1993,Kall:2009}. 
As for the majority of binary stars the observations are not sufficient to determine all parameters, for smoothing the light curves may be used “phenomenological fits”. Often were used trigonometric polynomials (=”restricted Fourier series”), following a pioneer work of \citep{Pick:1881} and other authors, see \citep{Par:1936} for a detailed historical review. \citep{And:2010,And:2012} proposed a method of phenomenological modeling of eclipsing variables (most effective for algols, but also applicable for EB and EW – type stars).

\section{"Simplified" Model}
\begin{figure}[t]
\begin{center}
\includegraphics[width=0.64\linewidth, bb=0 0 1359 950]{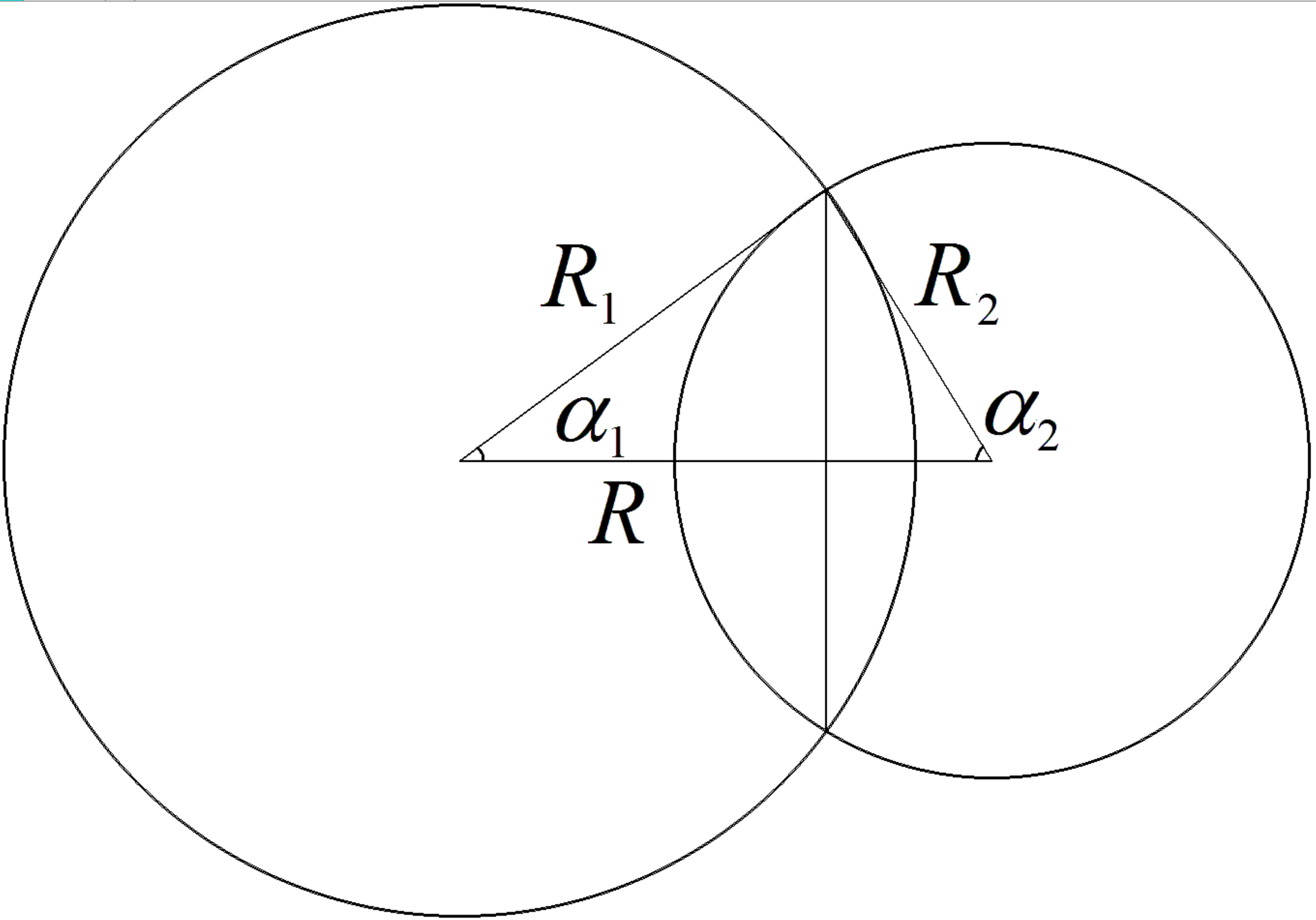}
\end{center}
\caption{\label{Fig1}Scheme of eclipsing binary system with spherical components}
\end{figure}

The simplest model is based on the following main assumptions: the stars are spherically symmetric (this is physically reliable for detached stars with components being deeply inside their Roche lobes); the surface brightness distribution is uniform. This challenges the limb darkening law, but is often used for teaching students because of simplicity of the mathematical expressions, e.g.  \citep{And:1991}. Similar simplified model of an eclipsing binary star is also presented by Dan Bruton (http://www.physics.sfasu.edu/astro/ebstar/ebstar.html). 
The scheme is shown in Fig.\ref{Fig1}. The parameters are $L_1, $ $L_2$ (proportional to luminosities), radii $R_1$, $R_2$, distance R between the projections of centers to the celestial sphere. 

\begin{figure}[t]
\begin{center}

\includegraphics[width=0.64\linewidth, bb=0 0 1956 1399]{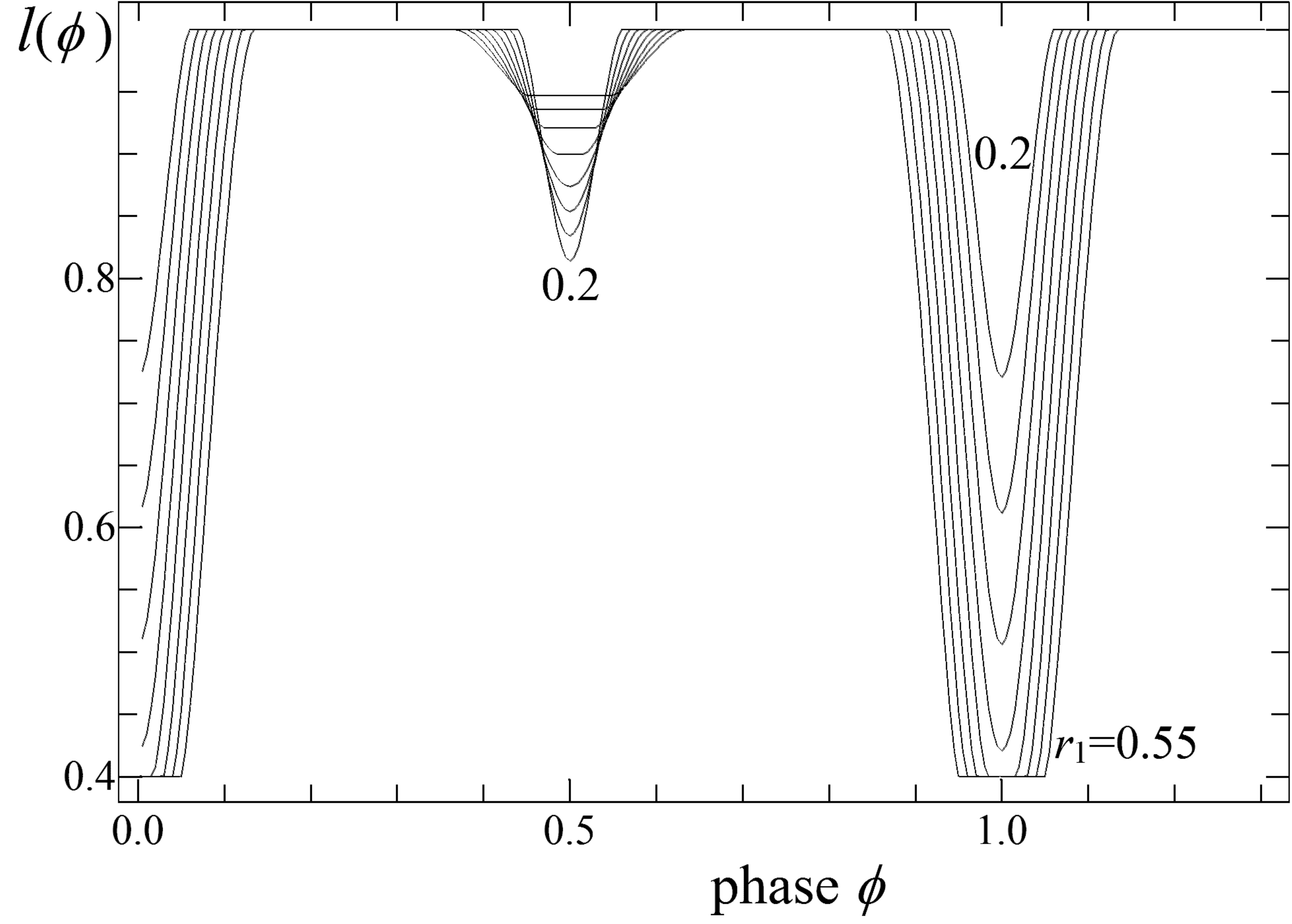}
\end{center}
\caption{\label{Fig2}A set of theoretical light curves for the "simplified" model generated for $R_1$ in a range from 0.2 to 0.55 with a step of 0.05 for fixed values of other parameters listed in the text}
\end{figure}

The square of the eclipsed segment is $S=S_1+S_2$
\begin{gather}
S_1=R_1^2(\alpha_1-\sin\alpha_1 \cos\alpha_1),               \label{eq1}\\
S_1=R_2^2(\alpha_2-\sin\alpha_2 \cos\alpha_2),                \label{eq2}
\end{gather}
where the angles $a_1, $ $a_2$ may be determined from the cosine theorem:
\begin{gather}
\cos\alpha_1=\frac{R^2+R_1^2-R_2^2}{2R_1R}=\frac{R^2+\eta}{2R_1R}, \label{eq3}\\
\cos\alpha_2=\frac{R^2+R_2^2-R_1^2}{2R_1R}=\frac{R^2-\eta}{2R_2R}, \label{eq4}
\end{gather}
where obviously  $\eta=R_1^2-R_2^2$ and $|R_1-R_2| \leq R \leq |R_1+R_2|$. The total flux is $L=L_1+L_2$, if $R\ge R_1+R_2$ 
(i. e. both stars are visible, $S=0$). For $R\le R_1+R_2$, $S=\pi R_2^2$ (assuming that $R_2\le R_1$). Generally, $L = L_1+L_2 – S/\pi R_j^2$, where $j$ is the number of star which is behind another, i. e. $j=1, $ if $\cos2\pi\phi\le0, $ and $j=2$, if $\cos2\pi\phi\ge0.$ Here $\phi$ is phase ($\phi=0$) corresponds to a full eclipse, independently on which star has larger brightness). For scaling purposes, a dimensionless variable $l(\phi)=L(\phi)/(L_1+L_2)$ is usually introduced. 
For tests, we used a light curve generated for the following parameters: $R_1=0.3, $ $R_2=0.2, $ $L_1= 0.4, $ $L_2=0.6$ and $i=80^\circ. $ The phases were computed with a step of 0.02. This light curve as well as other generated for a set of values of $R_1$ is shown in Fig.\ref{Fig2}. 
As a test function we have used:
\begin{equation}
F=\sum_{i=1}^n\frac{(x_i-\alpha x_c(\phi_i))^2}{\sigma_i^2},       \label{eq5}
\end{equation}
where  $x_i$(or $l_i$) are values of the signal at phases $\phi_i$ with a corresponding accuracy estimate $\sigma_i, $ and $x_c$ are theoretical values computed for a given trial set of $m$ parameters. 
For normally distributed errors and absence of systematic differences between the observations and theoretical values, the parameter $F$ is a random variable with $X_{n-m}^2$ a  statistical distribution \citep{Ander:2003,Cher:1993}. For the analysis carried out in this work, we used a simplified model with $\sigma_i=1$. This assumption does not challenge the basic properties of the test function. 
The scaling parameter is sometimes determined as $x(0.75)/x_c(0.75)$, i. e. at a phase where both components are visible, and the flux (intensity) has its theoretical maximum (in the “no spots” model). To improve statistical accuracy, it may be recommended to use a scaling parameter computed for all real observations:
\begin{equation}
 \alpha=\frac{\sum_{i=1}^n\frac{x_i}{\sigma_i^2}}{\sum_{i=1}^n\frac{x_c(\phi_i)}{\sigma_i^2}},   \label{eq6}
\end{equation}
This corresponds to a least squares estimate of a scaling parameter. I. e the model value of the out--of--eclipse intensity $L=L_1+L_2$ may be theoretically an any positive number, and these parameters may be "independent". By introducing $l_1=L_1/L$ and $l_2=L_2/L, $ we get an obvious relation $l_2=1-l_1, $ i. e. one parameter. For $L, $  sometimes are used values at the observed light curve at the phase 0.75 (i. e when both stars are to be visible so maximal light). We prefer instead to use all the data with scaling as in Eq.(\ref{eq6}). 
Even in our simplified model, the number of parameters is still large (4). At Fig.\ref{Fig4}, the lines of equal levels of $F$ are shown. One may see that the zones of small values are elongated and inclined showing a high correlation between estimates of 2 parameters. In fact this correlation is present for other pairs of parameters. This means that there may be relatively large regions in the multi–parameter space which produce theoretical light curves of nearly equal coincidence with observations. 

\begin{figure}[t]
\begin{center}
\includegraphics[width=0.64\linewidth, bb=0 0 928 580]{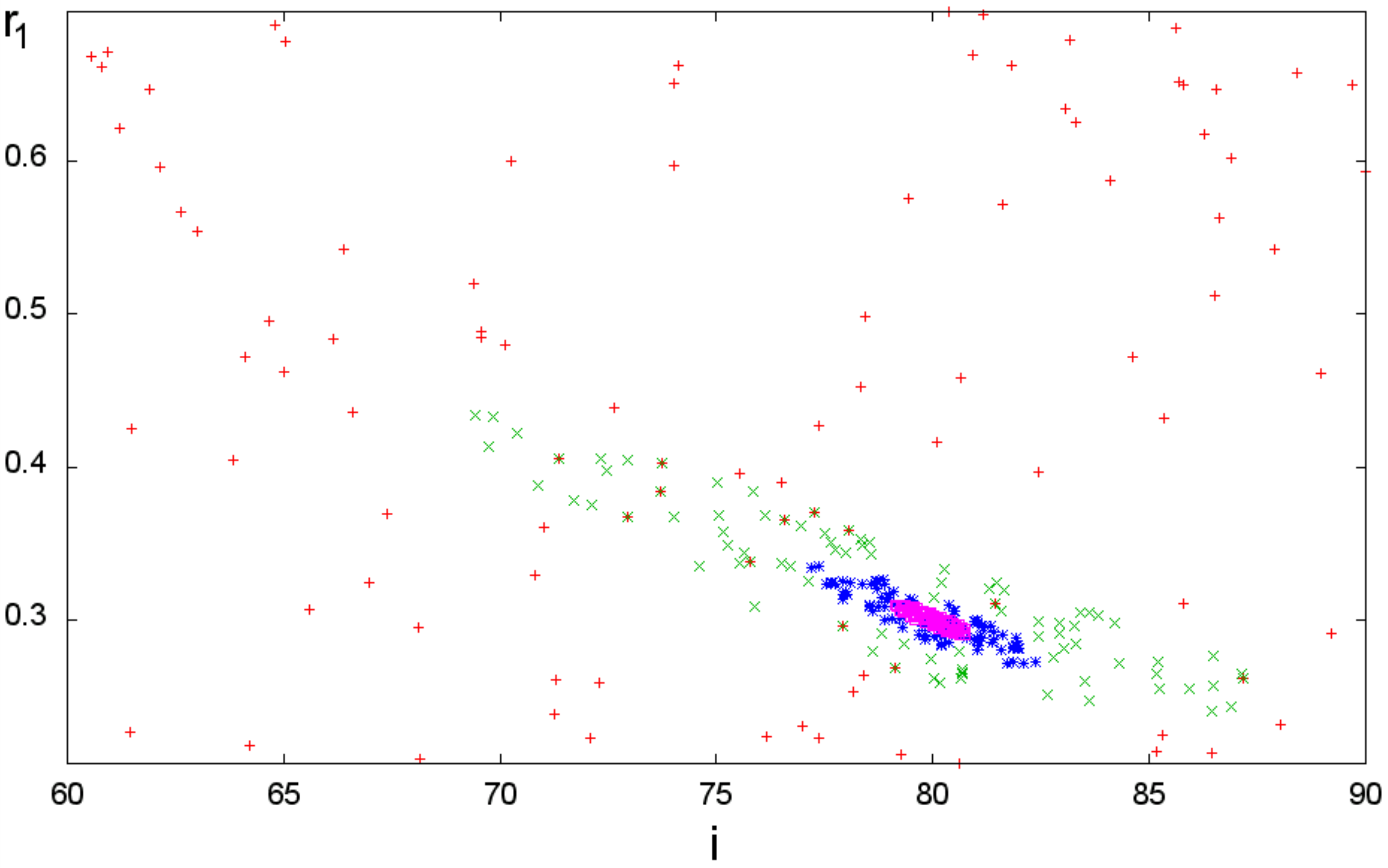}
\end{center}
\caption{\label{Fig3}Best 100 points after $10^2, $ $10^3, $ $10^4, $ $10^5$ trial computations, respectively.}
\end{figure}

\begin{figure}[b]
\begin{center}
\includegraphics[width=0.64\linewidth, bb=0 0 792 456]{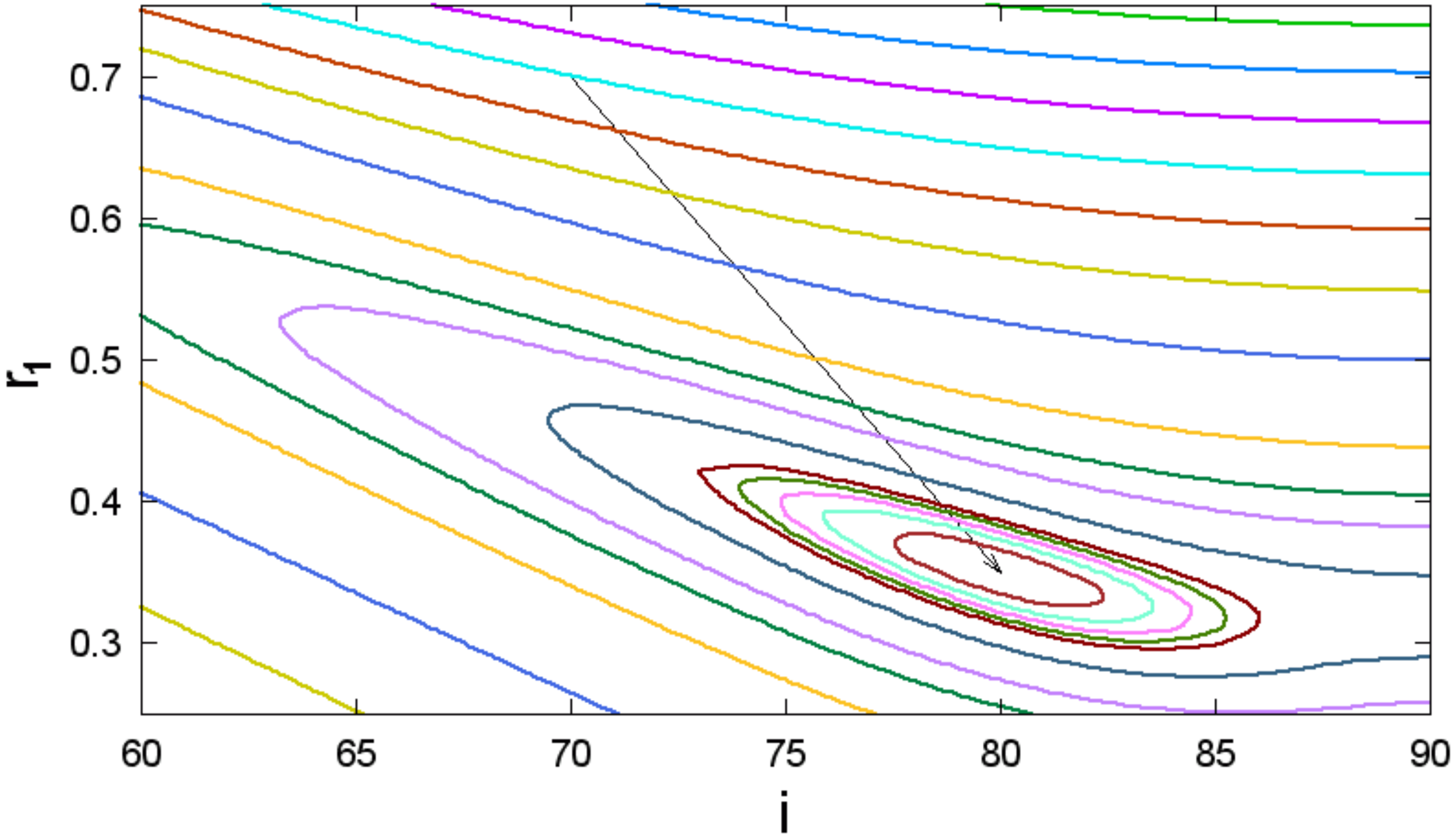}
\end{center}
\caption{\label{Fig4}Lines of equal values of the test function $F$ for fixed values of other parameters. The arrow shows position of the “true” parameters used to generate the signal. }
\end{figure}

In the software by \citep{Zol:2010}, the Monte-Carlo method is used, and at each trial computation of the light curve, the random parameters are used in a corresponding range: 
$C_k=C_{k, min}+(C_{k, min}-C_{k, min}){\rm rand},$                   
 where rand is an uniformly distributed random value. 
Then one may plot “parameter – parameter” diagrams for “best” points after a number of $N$ trial computations. The “best” means sorting of sets of the parameters according to the values of the test function $F$. 
Initially, the points are distributed uniformly. With an increasing $N$, “better” (with smaller $F$) point concentrate to a minimum. There may be some local minima, if the number of parameters will be larger (e.g. spot(s) present in the atmosphere(s) of component(s)). 
 We had made computations for an artificial function of $m(=1,2,3)$ variables \citep{And:2013a}. The minimal value $\delta$ (as a true value was set to zero), which was obtained using $N$ trial computations in the Monte-Carlo method is statistically proportional to
\begin{equation}
 \delta\propto N^{-2/m},                 \label{eq8}
\end{equation}
 i.e. the number of computations $N\propto\delta^{-m/2}$ drastically increases with both an increasing accuracy and number of parameters. 
For our simplified model, the numerical experiments statistically support this relation. Also, the distance between the “successful computations” (when the test function becomes smaller than all previous ones) $\Delta N\propto N$. Obviously, it is not realistic to make computations of the test function for billions times to get a set of statistically optimal parameters. 
In the “brute force” method, the test functions are computed using a grid in the $m$ – dimensional space, so the interval of each parameter is divided by $n_i$ points. The number of computations is $N=n_1n_2…... n_m$ should be still large. Either the Monte Carlo method, or the “brute force” one allow to determine positions of the possible local extrema in an addition to the global one. 
However, if the preliminary position is determined, one should use faster methods to reach the minimum. Classically, there may be used the method of the “steepest descent” (also called the "gradient descent"), where the new set of parameters may be determined as
\begin{equation}
 C_{k+1, i}=C_{k, i}-\lambda h_{k, i},              \label{eq9}
\end{equation}
where  $C_{k, i}$ is the estimated value of the coefficient $C_i$ at $k$-th iteration, $h_{k, i}$ – proposed vector of direction for the coefficient $C_i$, and $\lambda$ is a parameter. Typically one may use one of the methods for one--dimensional minimization \citep{Pres:2007, Kor:1968}, determine a next set of the parameters $C_{k, i}$, recompute a new vector $h_{k, i}$ and again minimize $\lambda. $ In the method of the steepest descent, one may use a $h_{k, i}=\partial F/\partial C_i$ gradient as a simplest approximation to this vector. Another approach (Newton-Raphson) is to redefine a function $F(\lambda)=F(C_i, i=1... m)$, compute the root of equation $\partial F/\partial \lambda =0, $ and then to use a parabolic approximation to this function. Thus
\begin{equation}
\lambda =(\partial F/\partial \lambda)/(\partial^2 F/\partial \lambda^2).               \label{eq10}
 \end{equation}
 
There may be some modifications of the method based on a decrease of $\lambda$, which may be recommended, if the shape of the function significantly differs from a parabola. 
In the method of “conjugated gradients”, the function is approximated by a second-order polynomial. Finally it is usually recommended to use the \citep{Marq:1963} algorithm. We tested this algorithm with a combination of the “steepest descent” (when the determinant of the Hessian matrix is negative) and “conjugated gradients” (if positive), which both are efficient for a complex behavior of the test function. 

\section{Phenomenological Modeling}
Besides physical modeling of binary stars, there are methods, which could be introduced as "phenomenological" ones. In other words, we apply approximations with some phenomenological parameters, which have no direct relation to physical parameters - masses, luminocities, radii etc. 
The most often used are algebraic polynomial approximations, included in the majority of computer programs (e.g. electronic tables like Microsoft Office Excel, Libre/Open Office Calc, GNUmeric etc.). For periodic processes, one can use a trigonometric polynomial (also called "restricted sum of Fourier series"
\begin{gather}
x_c(\phi, s)=C_1+\sum_{j=1}^s (C_{2j}\cos(2j\pi\phi)+C_{2j+1}\sin(2j\pi\phi))\label{11}\\ 
~~~~~~~~~=C_1+\sum_{j=1}^s R_j\cos(2j\pi(\phi-\phi_j))\nonumber 
\end{gather}
The upper Equation is used for determination (using the Least Squares method) of the parameters $C_\alpha, $ $\alpha=1..m, $ where the number of parameters is $m=1+2s$, where the lower converts the pairs of the coefficients $C_{2j+1}, $ $C_{2j+1}$ for each $(j-1)$-th harmonic according to usual relations
\begin{gather}
C_{2j}=R_j\cos(2j\pi\phi_0)\nonumber\\ 
C_{2j+1}=R_j\sin(2j\pi\phi_0)\label{12}\\
R_j=(C_{2j}^2+C_{2j+1}^2)^{1/2}\nonumber\\ 
\phi_j={\rm atan}(C_{2j+1}/C_{2j})/2\pi+0.25(1-{\rm sign}(C_{2j})) \nonumber 
\end{gather}

\begin{figure}
\begin{center}
\includegraphics[width=1.3\linewidth, bb=0 0 2221 1216]{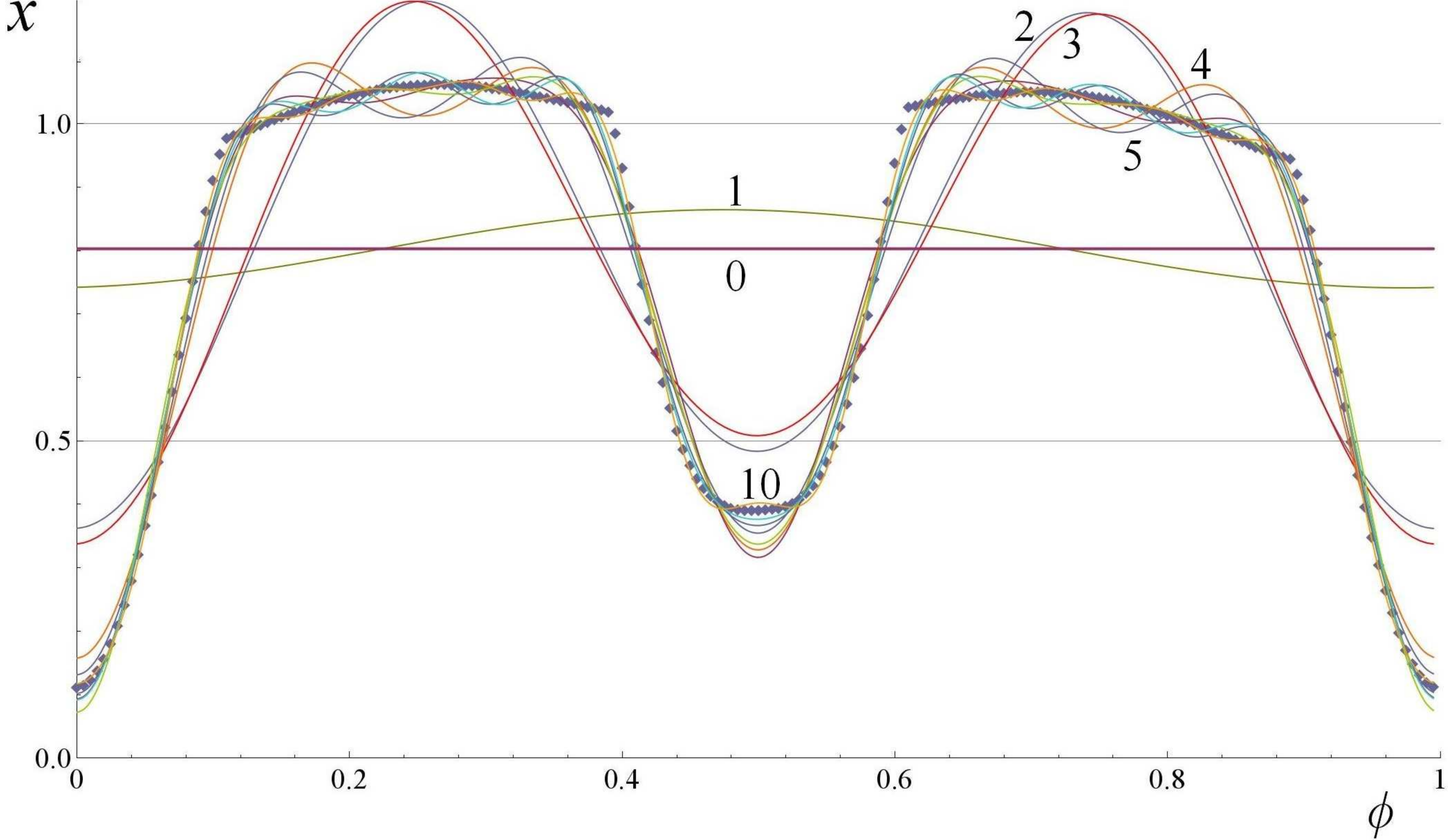}
\end{center}
\caption{\label{Fig5}Trigonometrical polynomial approximations of the phenomenological light curve. The degree $s$ is shown by numbers near corresponding curves. }
\end{figure}

\begin{figure}
\begin{center}
\includegraphics[width=1.3\linewidth, bb=0 0 1367 677]{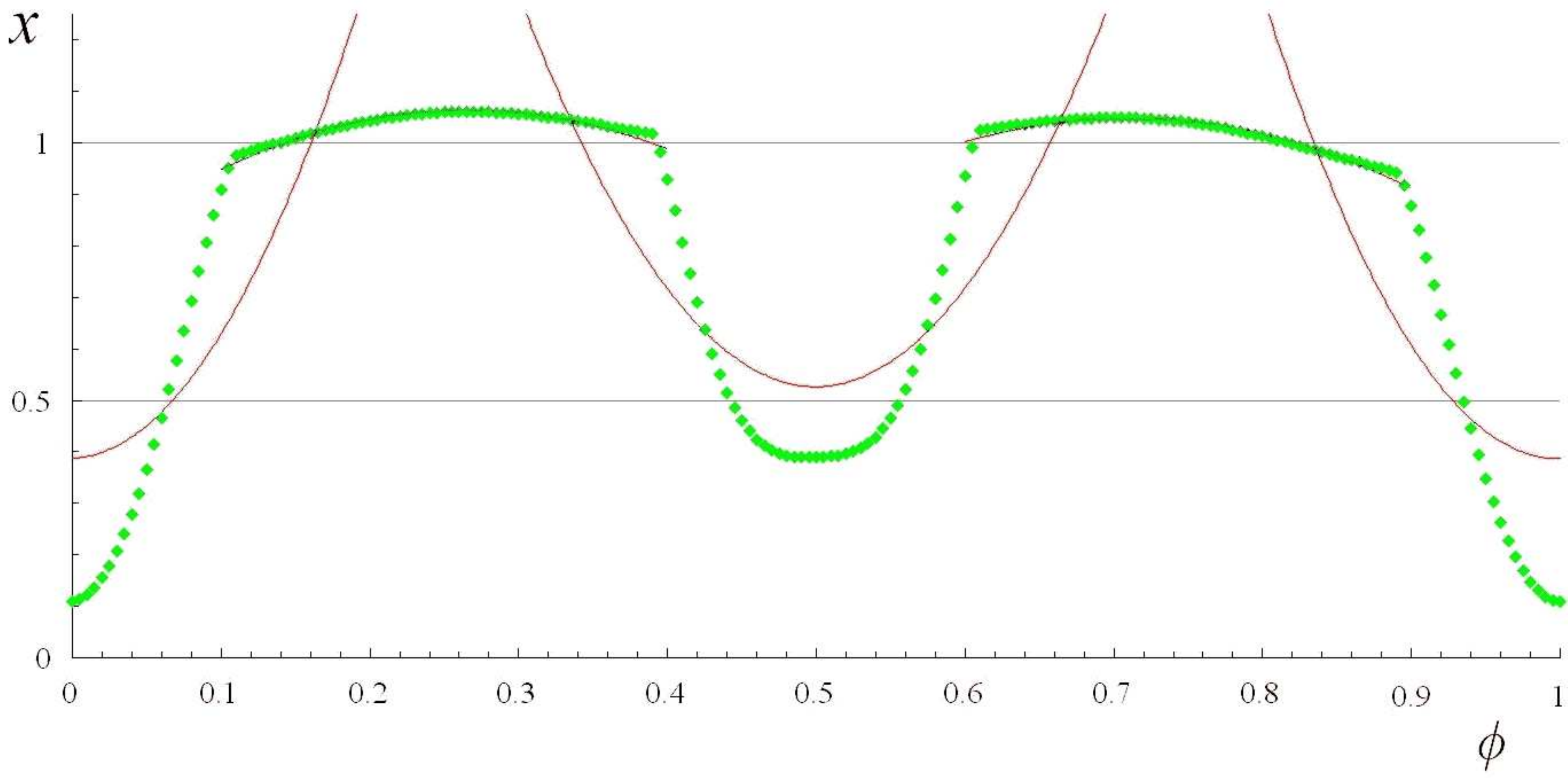}
\end{center}
\caption{The model light curve and its approximation by parabola at the intervals of phases centered on mainima and maxima, as proposed by \citep{Pap:2014}\label{Fig6}}
\end{figure}

Here $j=1..j_{max}$, $j_{max}=n/2$ for even $n$ and $j_{max}=(n-1)/2$ for odd $n. $
Using the Least Squares algorithm, it is possible to determine parameters $C_\alpha$ even for irregularly spaced data e.g. \citep{And:1994}. 
{\em Only} under strong conditions $\phi_k=\phi_0+k/n$, where $k=0..n-1, $  $n$ is the number of observations, one may obtain simplified expressions for the "Discrete Fourier Transform" (DFT) as an extension of the original Fourier (1822) method:
\begin{gather}
C_{0}=\frac{1}{n}\sum_{k=0}^{n-1} x_k\nonumber\\
C_{2j}=\frac{2}{n}\sum_{k=0}^{n-1} x_k\cos(2j\pi k/n) \label{13}\\
C_{2j+1}=\frac{2}{n}\sum_{k=0}^{n-1} x_k\sin(2j\pi k/n)\nonumber
\end{gather}
If $j=n/2, $ then 
\begin{gather}
C_{n}=\frac{1}{n}\sum_{k=0}^{n-1} (-1)^kx_k, \label{14}\\
C_{n+1}=0, \nonumber
\end{gather}
For irregularly spaced data, there are at least 6 different modifications of the method, which are called themselves as "Fourier Transform", and give same correct results only under assumptions listed above for the DFT. For irregularly spaced data The links may be found in \citep{And:2003}. 

Theoretically, the degree of the trigonometric polynomial $s$ is infinite for continuous case (number of data $n\to\infty$) and should be $s=j_{max}={\rm int}(n/2), $ i. e. may be a large number. For this case, one will get an interpolating function. For lower degree $s<j_{max}, $ the function is smoothing, and one may use different criteria for choosing the statistically optimal value, e.g. the Fischer's criterion (or equivalent one based on the Beta--type distribution), the criterion of minimum of r. m. s. error estimate of the smoothing function (at the moments of observations; integrated over all interval; or at some specific value of the argument), or the maximum of the "signal--to--noise" ratio. 

However, these sums may show apparent waves (so--called Gibbs phenomenon). It may be illustrated in Fig.(\ref{Fig5}) for a sample function. One may see different approximations. With an increasing $s, $ the approximation $x_c(\phi, s)$ becomes closer (in a sense of the Least Squares), but the apparent waves are well pronounced at $m<<n. $ 

To decrease the number of parameters, \citep{And:2010, And:2012} proposed an approximation combined from a second--degree trigonometric polynomial and a local function modeling the shape of the eclipses:

\begin{gather}
 x_c(\phi)= C_1+C_2\cos(2\pi\phi)+C_3\sin(2\pi\phi)+\nonumber\\
~~~~~~~~~~~~~~~ +C_4\cos(4\pi\phi)+C_5\sin(4\pi\phi)+\\
~~~~~~~~~~~~~~~ +C_6H(\phi-\phi_0, C_8, \beta_1)+C_7H(\phi-\phi_0+0.5, C_8, \beta_2). \nonumber                                          \label{eq15}
\end{gather}

\begin{equation}
 H(\phi, C_8, \beta)=\left\{
 \begin{array}{ll}
 V(z)=(1-|z|^\beta)^{3/2}, & {\rm ~~if~} |z|<1\\
  0, & {\rm ~~if~ } |z|\geq 1
 \end{array} 
 \right., 
 \label{eq16}
\end{equation}
where $z=2\phi/D$, $\phi=E-{\rm int}(E+0, 5)$ -- phase, $E=(t-T_0)/P$ -- (non-integer) cycle number, $t$ -- time, $T_0$ -- initial epoch, $P$ -- period, $D$ -- full duration of minimum in units of $P.$

\begin{figure}[t]
\begin{center}
\includegraphics[width=0.56\linewidth, bb=0 0 1999 1099]{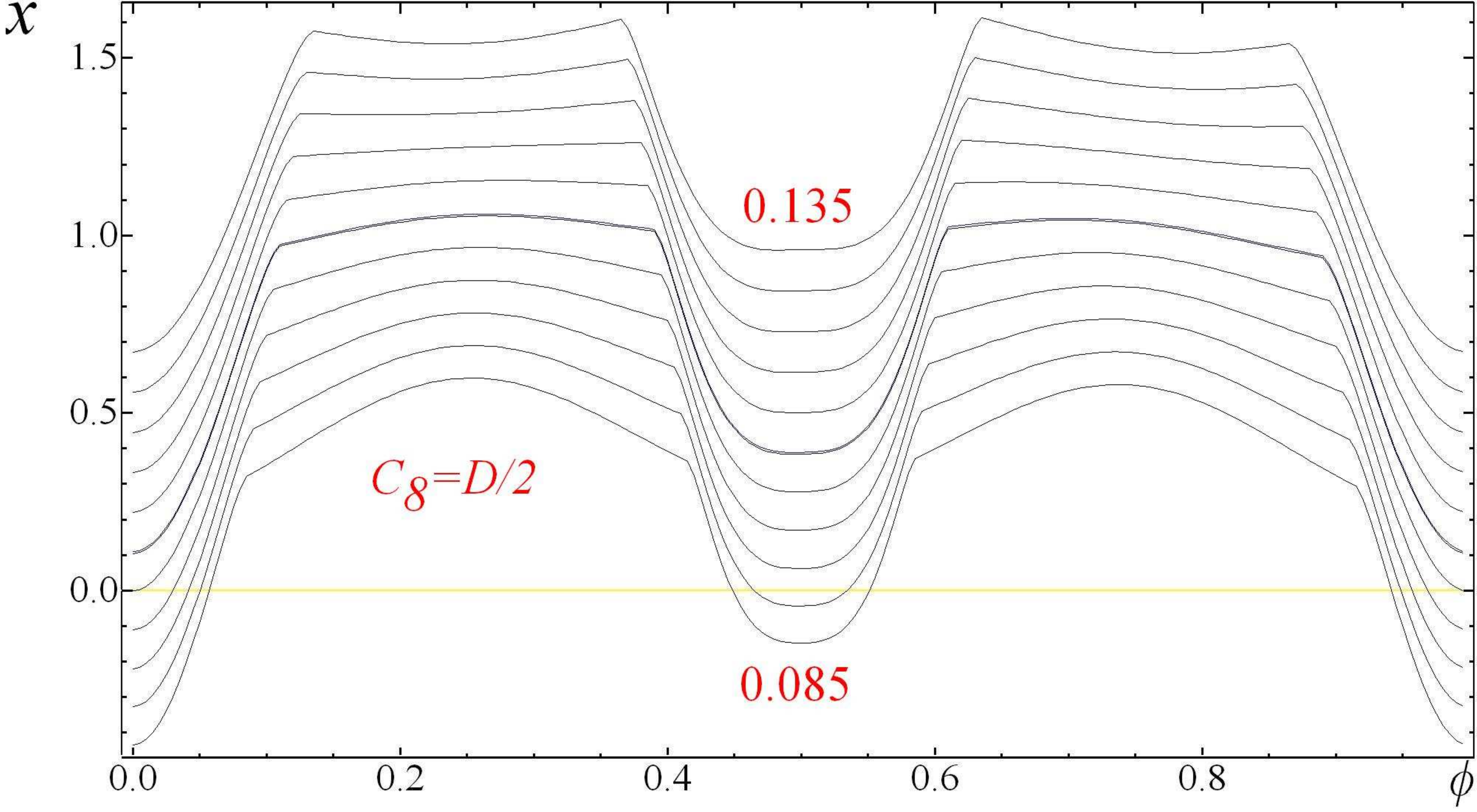}~
\includegraphics[width=0.56\linewidth, bb=0 0 1999 1099]{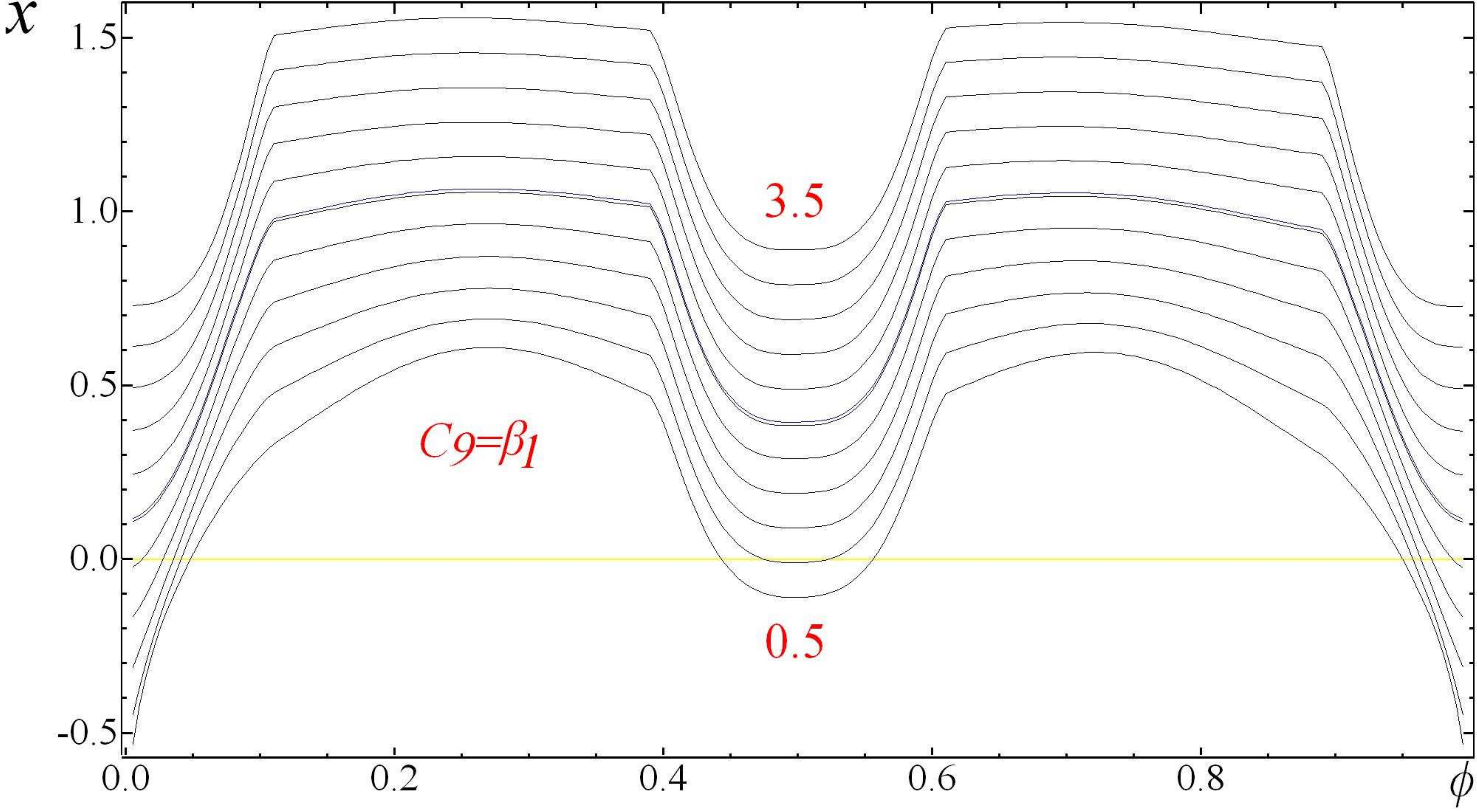}
\end{center}
\caption{\label{Fig7}Dependencies of the light curves (intensity vs. phase) on the parameters $C_8=D/2$ (left) and $C_9=\beta_1$ (right). The relative shift in intensity between subsequent curves is 0.1. The thick line shows a best fit curve}
\end{figure}

\begin{figure}[t]
\begin{center}
\includegraphics[width=0.56\linewidth, bb=0 0 1999 1099]{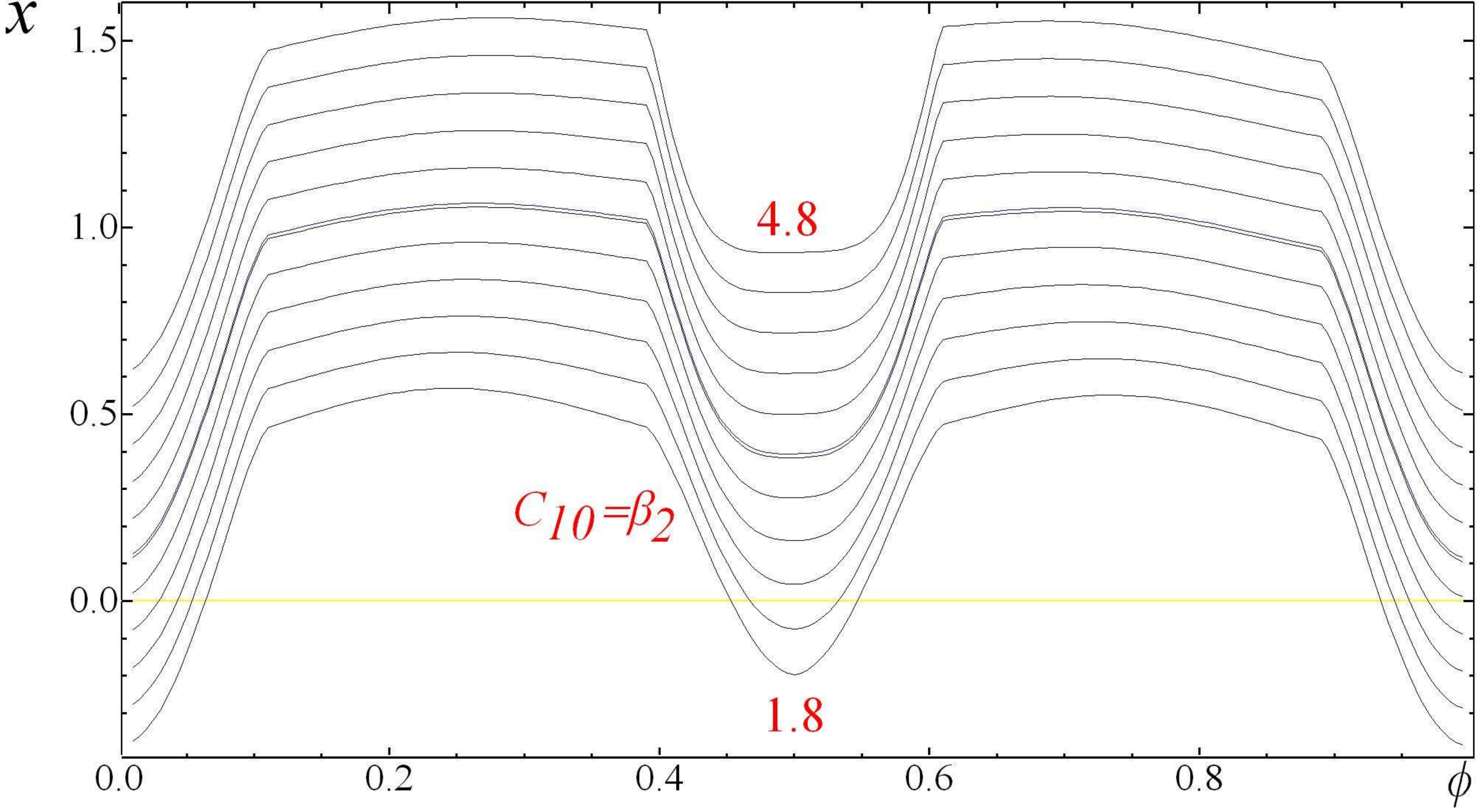}~
\includegraphics[width=0.56\linewidth, bb=0 0 1999 1099]{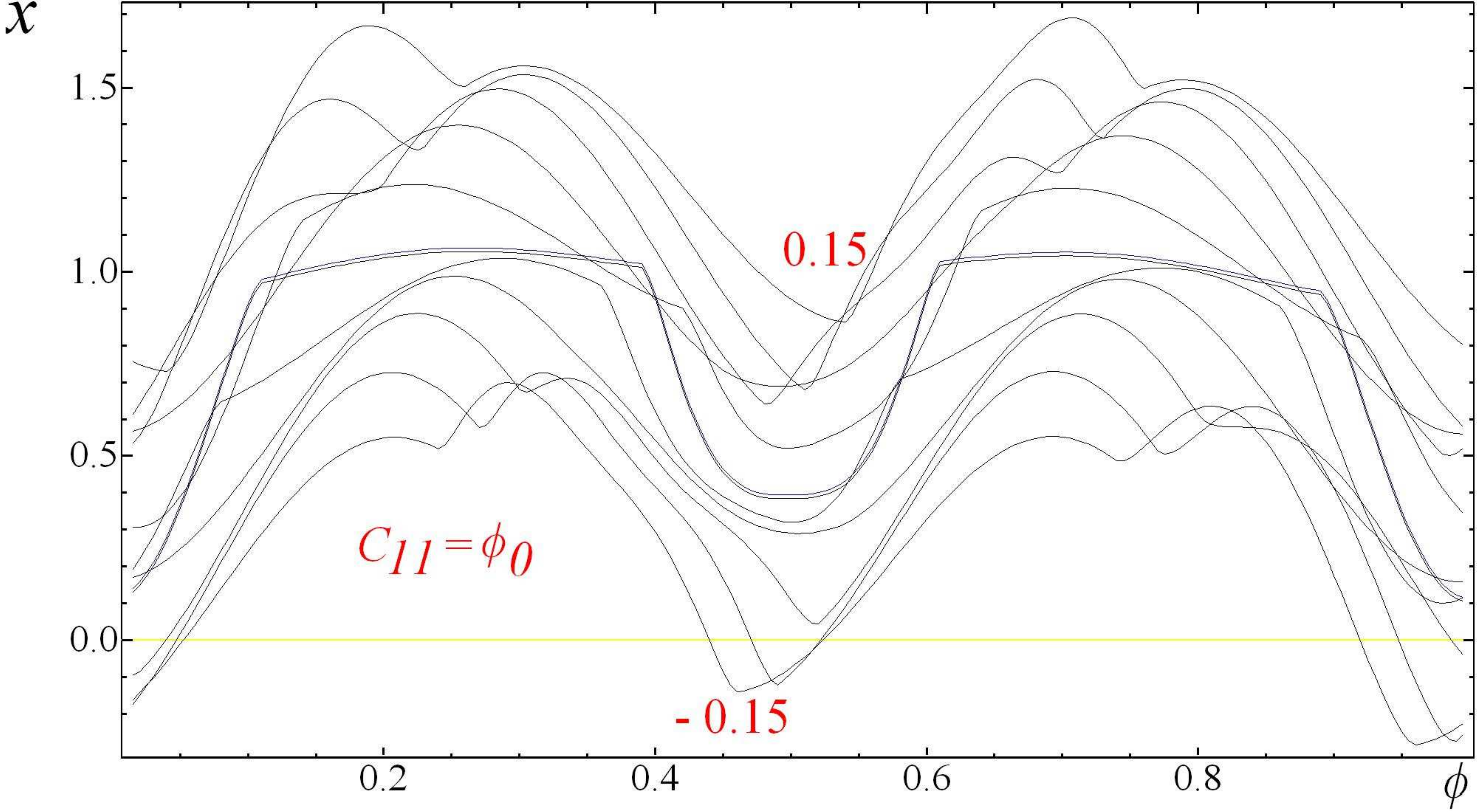}
\end{center}
\caption{\label{Fig8}Dependencies of the light curves on the parameters $C_{10}=\beta_2$ (left) and $C_{11}=\phi_0$ (right). }
\end{figure}

\citep{Pap:2014} made a statistical study of a sample of eclipsing binaries. They have used an oversimplified approximations of the light curves, approximating the my a parabolic fit over overlapping intervals 
$[-0.2,+0.2],$ 
$[0.1,0.4],$ 
$[0.3,0.7],$ 
$[0.6,0.9],$ 
$[0.8,1.2].$ 
Obviously, the first and last interval correspond to the same observations. In fig(\ref{Fig6}) we show their fit to our sample light curve. One may see a relatively good approximation of the out-of-eclipse part of the light curve, and a bad approximation of the zone of minimum. A better coincidence of the fit near minima may be expected for EW--type stars, whereas for EA--type stars our NAV algorithm produces significantly better approximation for all phases. 

To illustrate the dependence of the "best fit" light curves on the "non-linear" parameters 
$C_8..C_{11}, $ 
we show corresponding approximations in Fig.(\ref{Fig7}) and Fig.(\ref{Fig8}). The thick line in the middle of each figures corresponds to the curve for the sample parameters 1, --0.04, 0.01, --0.05, 0.01, --0.8, --0.6, 0.11, 2, 3.3, 0 for $C_1..C_{11},$ respectively. 

One may see the significant variations of the shape of the curve and, for each real observations, the best fit solution is expected to be unique. As in previous cases, the solution may be determined using different methods.

We developed the software realizing various methods for study of variable stars. The results of this study will be used in the frame of the projects "Ukrainian Virtual Observatory” (UkrVO) \citep{Vav:2012} and “Inter-Longitude Astronomy“ \citep{And:2010ea}.

\end{document}